\documentclass[10pt]{article}
\usepackage{epsfig,amstex}

\begin{document}


\title{Finite-temperature evaluation of the Fermi density operator}

\author{Florian Gagel \\
        Institut f\"ur Physik, \\ 
        Technische Universit\"at Chemnitz-Zwickau, \\
        D-09107 Chemnitz, Germany \\
        Email: f.gagel@@physik.tu-chemnitz.de
        }

\date{\today}
\maketitle

\begin{abstract} \noindent
 A rational expansion of the Fermi density operator
 is proposed.
 This approach allows to 
 calculate efficiently physical properties of
 fermionic systems at finite temperatures 
 without solving an eigenvalue problem.
 Using $N$ evaluations of the Green's function, the
 Fermi density operator can be approximated, 
 subject to a given precision,
 in the energy interval $[-\beta,\infty]$ with $ \beta \propto N$.
 The presented method may become especially useful for 
 electronic structure calculations involving the calculation of 
 charge densities, but may also find other applications in 
 e.g.~signal processing and numerical linear algebra. 
\end{abstract}

\vfill


\vspace{1cm}
\noindent
{\bf Classification:}

\vspace{1cm} \noindent
\begin{tabular}{ll}
65D20 & Computation of special functions, construction of tables \\
81-08 & Computational methods \\
81Q05 & Closed and approximate solutions to the Schrodinger, \\
      &   Dirac, Klein-Gordon and other quantum-mechanical
               equations \\
\end{tabular}

\clearpage


\section{INTRODUCTION}
Quantum systems are most generally described in terms
of their density operator $\bf \rho$. Once $\bf \rho$ is
known, the  expectation values of physical quantities
are  obtained as
\begin{equation}
  \langle A \rangle \; = \; \text{Tr} ({\bf \rho} {\bf A})
\end{equation}
where ${\bf A}$ is the associated operator of the
quantity under consideration. For instance, for the
calculation of the charge
density, $\bf A$ becomes a projector
and the charge density is simply given by the 
diagonal elements of $\rho$ 
in the site representation. In the following we consider
fermionic systems in the grand-canonical ensemble where
\begin{equation} \label{eq:rhodef}
  \rho(\mu,T) \;=\; f( \frac{ {\bf H} - \mu }{ kT } ),
\end{equation}
with $\bf H$, $T$ and $\mu$ 
being the Hamiltonian, the temperature and the chemical potential
respectively, 
and 
\begin{equation} \label{eq:fermi}
  f(x) = \frac{1}{1+e^x}
\end{equation}
being the Fermi function.
The Fermi function has been studied extensively,
and effective approximation schemes 
for the case  of scalar arguments, 
e.g.~the Sommerfeld expansion,
have been developed \cite{ashcroft}.

\noindent
However, for the calculation of e.g.~$\bf \rho$, one
is faced with $f$ applied to operators.
For large scale applications, it cannot be
be switched into the eigen representation of $H$
in order to evaluate Eq.~\ref{eq:rhodef}, since in general
full diagonalization of $\bf H$ is practically impossible. 
Because only polynomial and fractional functions of operators can 
be evaluated, corresponding decompositions
of $f(x)$ are highly desirable. 
A recent approach is due to Goedecker \cite{goedeck} 
who proposed to
use systematically complex line integrals over 
the Green's function.
In the following a fractional expansion is presented which does not
depend on  the calculation of line integrals. 
It will be shown 
that physical quantities like the
charge density, which is at the base of many methods
in electronic structure calculations, can be obtained effectively
{\em without} solving an eigenvalue problem, necessitating only
evaluations of the Green's function at selected points.
While the method is {\em a priory} constructed for
finite temperatures, it is also well adapted to approximate 
charge densities at zero temperature since the range 
of the approximation can be arbitrarily extended towards lower
temperatures.

\section{The fractional expansion}
It is well known, that the Matsubara expansion of 
the Fermi function,
\begin{equation}
  f(x) = 1/2 - 2 \sum_{m=0}^{\infty} \frac{x}{x^2+[(2m+1)\pi]^2}
\end{equation} 
shows very poor convergence properties when 
truncated to degree $m=N$. Although being the
exact fractional series of $f(x)$, the Matsubara expansion 
is therefore not suited for numerical applications.

\noindent
Let us consider the function
\begin{equation} \label{eq:fa}
  f_\alpha(x) \,:=\, f(x-\alpha) f(-x-\alpha) \,=\, 
                \frac{e^\alpha}{2[ \cosh(\alpha) + \cosh(x)]},  
\end{equation}
which is depicted in Fig.~1 for $\alpha=20$.
It is readily seen that for sufficiently high $\alpha>0$, 
$f_\alpha(x+\alpha)$ will approximate $f(x)$ for
all $x > -\alpha$ subject to a given precision. 
We now truncate the series in the denominator,
\begin{equation} \label{eq:frac}
  f_\alpha(x) \approx g_N(x; \alpha) := \frac{e^\alpha}
                                        {2\,p_N(x;\alpha)},
\end{equation}
where 
\begin{equation}
  p_N(x;\alpha) = \cosh{(\alpha)} + \sum_{j=0}^{N} \frac{x^{2j}}{(2j)!}.
\end{equation} 
It is readily seen than $p_N(x;\alpha)$ has no real zeros.
For the fractional expansion of Eq.~\ref{eq:frac}, we need all zeros 
$z_\nu,\; \nu=1,\dots,N$ of $q(z) := p_N(x;\alpha)|_{x^2=z}$
(see \cite{drop}). 
For this purpose we define 
\begin{equation} \label{eq:ydef} 
  y_1 = 1 + \cosh(\alpha); \;\, y_i(z) = \frac{z^{i-1}}
                                      {[2(i-1)]!},\;\; i=2,\dots,N-1. 
\end{equation}
Then, it can be seen that 
\[ {\bf y}(z) \equiv (y_1(z),...,y_N(z))^T\] 
satisfies a matrix equation
$ {\bf A} \, {\bf y}(z_{\nu}) \;=\; z_{\nu} \, 
                                     {\bf y}(z_{\nu})$,
with the $N \times N$ matrix ${\bf A} \equiv (a_{i,j})$,
\begin{equation}
  a_{i,j} = \left\{ 
                \begin{array}{ll}  \label{eq:Adef} 
                 2+2\cosh(\alpha)  & \text{if } (i,j) = (1,2); \\
                 2l (2l-1)         & \text{if } (i,j) = (l,l+1), \;
                                      l=2,\dots,N-1; \\
                  2N\,(1-2N)         & \text{if } i=N; \\
                        0          & \text{else.}
                 \end{array}
            \right.
\end{equation}
One easily shows that the $z_{\nu}$ are given by the eigenvalues
of the matrix ${\bf A}$. 
It is well known that the zeros of a given
polynomial can be obtained from an eigenvalue problem
for a related Hessenberg matrix.
Goedecker\cite{siam} already has proposed to use this fact
for the numerical evaluation of all zeros of a polynomial as eigenvalues. 
The usual scheme corresponds to the implicit choice
of $y_i = z^{i-1}$; here the point
is to avoid the explicit use of any factorial by using
Eq.~\ref{eq:ydef} leading to Eq.~\ref{eq:Adef}. 
The $z_{\nu}$ can be obtained as 
eigenvalues with e.g.~QR-rotations
in a numerically stable way; using 
standard numerical libraries, $N=40$ still
yields accurate results, and enhanced precision calculations 
readily allow for larger $N$. However, $N=30$ will be already 
sufficient for many applications as will be shown.
In the following we stick to even $N$.
For convenience,
we also chose $N$ and
$\alpha$ so that no duplicate zeros $z_{\nu}$ are obtained.  
From the zeros we obtain the $2N$-zeros $x_\nu$ of $p_N(x;\alpha)$
as $\pm \sqrt{z_{\nu}}$. 
The zeros $z_{\nu;\alpha}$ do behave well, 
an example is plotted in Fig.~2(i). 
An important although trivial remark must be made on the evaluation of
polynomials such as $q(z)$ and $q'(z)$, which is to 
be done numerically using a Horner-like scheme in order
to avoid any explicit use of the factorial
for the reason of limited numerical precision, 
as e.g.~$40!$ already 
is a number with 39 decimal digits. 
Denoting 
\[ \gamma_{\nu}=\frac{e^\alpha}{2 \,q'(x_{\nu}^2) \,x_{\nu}},\] 
we now may write down the fractional decomposition
\begin{equation} \label{eq:fractexp0}
 f_\alpha(x) \approx  g_N(x;\alpha) \; = \;
      \sum_{\nu=1}^{2N} \frac{\gamma_{\nu}}
                              {x-x_{\nu}}.
\end{equation}
As shown in Fig.~2(ii), the coefficients 
$\gamma_{\nu}$ also behave well. 
The approximation Eq.~\ref{eq:fractexp0} converges rapidly.
Choosing e.g.~$N=32$ and $\alpha=26$, the error in
approximating $f_\alpha(x)$ is less than
$10^{-9}$ for all real $x$.
We have considered the symmetric function $f_\alpha$, 
since we may now exploit the {\em local symmetry} of $f_\alpha(x)$
about the points $x = \pm \alpha$ where $f_\alpha(x)=0.5$. 
We can approximate
successively the Fermi function as sum of shifted
functions $ g_N(x;\alpha) $,
\begin{equation} \label{eq:fractexp}
 f(x) \approx g_N(x+\alpha;\alpha) + g_N(x+3\alpha;\alpha) + ... + 
                                    g_N(x+(2M-1)\alpha;\alpha)
\end{equation}
in the range $[-(2M-1)\alpha, \infty]$. This is visualized
in Fig.~3 for $N=32$ and $\alpha=26$, using
$2MN=192$ fractional terms.
For $x \rightarrow \infty$, the approximation Eq.~\ref{eq:fractexp}
vanishes like $x^{-2N}$ compared to
exponential decay of the Fermi function, resulting naturally
in a good approximation. For negative $x$,
the validity range of the 
approximation Eq.~\ref{eq:fractexp}
may be increased by choosing a higher $M$, i.e.,~by successively 
adding shifted realizations of  $g_N(x+(2m-1)\alpha;\alpha)$.

\noindent
We note, that the function $f_{(\xi \alpha)}(\xi x)$ (see Eq.~\ref{eq:fa})
represents for sufficiently large $\xi$ a
nearly perfect
projector on the subspace $x \in [-\alpha, \alpha]$. The presented 
rational expansion may therefore also find applications
in other fields than physics, especially when applied to 
operators.

\section{Application to operators}
The main interest of the approximation Eq.~\ref{eq:fractexp} 
lies in its generalization
as operator equation, replacing $x$ by some Hamiltonian ${\bf H}$.
Then, the Fermi density operator 
\begin{eqnarray}    \label{eq:fermiop} 
  {\bf \rho}(kT) &=& \frac{1} 
                 {{\bf 1} + \exp( \frac{ {\bf H} - \mu}
                                         {k T} )} \\   \nonumber
      &\approx& k T \, \sum_{\nu=1}^{2N} \sum_{m=1}^{M}                     
                 \frac{\gamma_{\nu}}
                  {{\bf H} - \mu + k T [ (2\,m-1)\alpha - x_{\nu}] }
\end{eqnarray}
can be approximated efficiently with
$2MN$ evaluations of the Green's function.
One may furthermore benefit from the fact that the 
zeros $x_{\nu}$  as well as the
corresponding $\gamma_{\nu;\alpha}$ come in 
quartets $x_{\nu},x_{\nu}^*,-x_{\nu},-x_{\nu}^*$ if
the $z_\nu$ are distinct.

When applied to operators $\bf H$, the effect of the
approximation Eq.~\ref{eq:fermiop}  is
to cut off the contributions of states with eigenvalues 
smaller than  $\epsilon_l = \mu -  k T (2 M-1) \alpha$
(see also Fig.~3). 
This has no consequences
if the spectrum of $H$ is lower bounded with no eigenvalues
in this domain. There are certainly applications 
when this effect is 
wanted, e.~g.~when considering the contributions of sub-bands
separately. 

In the example of Fig.~3, $M=3$ may be 
too small for applications involving real metals, since the 
eigenvalue spectrum is covered down to -3.5 eV only at room temperature, 
and a higher $M$ may be needed.
However, $M=3$ and $N=32$ already is well adapted for
e.g.~two-dimensional electron gases in mesoscopic systems.
Assuming an Fermi level of about 15 meV, the Fermi density operator
can be approximated quite exactly at temperatures down to 1.5 Kelvin.

The following simple example demonstrates how the total charge density
can be obtained without solving an eigenvalue problem. 
Consider
the Hamiltonian
\[
   H = \sum_j \epsilon_j a_j^\dagger a_j- 
        \sum_{j,k} t_{j,k} (a_j^\dagger a_k + H.c.),
\]
where the on-site energies $\epsilon_j$ have
been chosen from a uniform random distribution $\epsilon_j \in ]3,5[$,
and the hopping amplitudes $t_{j,k}$ have been chosen as 1 for
$j,k \equiv (j_x,j_y)^T,(k_x,k_y)^T$ being nearest neighbors 
in the two-dimensional plane. This Hamiltonian describes 
free spinless electrons in a discrete two dimensional
space in presence of a random impurity potential.  
Hard wall boundary conditions have been assumed for a system
with $N_s = 15 \times 15$ sites, allowing conveniently for direct 
diagonalization. 
The chemical potential has been fixed
between the 25th and 26th smallest eigenvalue of $H$, 
$\mu = (\lambda_{25} + \lambda_{26}) /2$, i.e., the system
is in contact with a heath bath of constant chemical potential. 
We are interested in the total charge density as function of
temperature $T$ , especially in the limit of $T \rightarrow 0$. 

In our case, the considered  system is small enough to
calculate the charge
density ${n_j}$ at site $j$ exactly as
\[
   {n_j}  = \sum_{i=1}^{N_s} |u_j^{(i)}|^2 
                              f( \frac{\lambda_i-\mu}
                                          {kT}),
\]
were the $\bf u^{(i)}$ are the normalized eigenvectors, thus 
allowing for direct comparison.

We now approximate the
Fermi density operator ${\bf \rho}$ 
according to Eq.~\ref{eq:fermiop}.
As noted in the introduction,
the charge density is given in this case by  the diagonal elements 
of ${\bf \rho}$.
The results presented in Fig.~4 have been
obtained using $2MN=192$ evaluations of the Green's function,
thus necessitating the solution of linear systems of equations only. 
It is seen that the charge density is indeed 
very well approximated
in the domain where Eq.~\ref{eq:fractexp} 
approximates Eq.~\ref{eq:fermi} as discussed previously. 
The total charge density
at zero temperature (which is, of course, given by the number
of states with energy smaller than the chemical potential, i.e.,
25 in the present example),
is practically identical to the total
charge density at low temperatures.

\section{CONCLUSIONS}
A fractional approximation of the Fermi density operator
has been proposed and the necessary concepts
have been presented. This method becomes increasingly
appropriate for higher temperatures where the
numerical effort decreases. However, its 
range of convergence can be arbitrarily extended towards
lower temperatures.
It is expected
to be useful especially for large scale calculations 
at finite temperatures as e.g.~investigations of disordered systems, 
mesoscopic systems and electronic structure calculations in general.
 
\vspace{0.5cm}
\noindent
For valuable suggestions I am grateful to K.~Maschke.

{\bf \Large \noindent Figure captions}\\

\small
\noindent {\bf \normalsize Fig.~1:} 
     The symmetric function $f_\alpha(x)$ for $\alpha=20$. 
  
\vspace{0.2cm}

\noindent {\bf \normalsize Fig.~2:}
       Positions of the zeros $x_{\nu}$ (i) 
       and the fractional coefficients
       $\gamma_{\nu}$ (ii) in 
       the complex plane for the fractional expansion with
       $N=32$ and $\alpha=26$. 
   
\vspace{0.2cm}

\noindent {\bf \normalsize Fig.~3:}
          Fermi function $f(x)$ and the fractional expansion
          with $M=3,\,N=32,\,\alpha=26$. The dotted lines indicate
          the $M=3$ shifted addents. The error in approximating 
          $f(x)$ is less than $10^{-9}$ for $x \ge -135$.

\vspace{0.2cm}

\noindent {\bf \normalsize Fig.~4:}
         Total charge $n_{\text{tot}} = 
            \sum_{j=1}^{N_s} n_j$ as function
         of $\theta:=kT/(\mu-\lambda_{\text{min}})$, where
         $\lambda_{\text{min}}$ is the smallest eigenvalue 
           (see text). The error in approximating $n_{\text{tot}}$
           using Eq.~\ref{eq:fermiop} with $N=32$, $\alpha=18$ and $M=3$
          is less than $10^{-6}$. 

\begin{thebibliography}{99}
\bibitem{drop} The indices $\alpha$ and $N$ have been dropped for
               the reason of clarity.
\end{thebibliography}
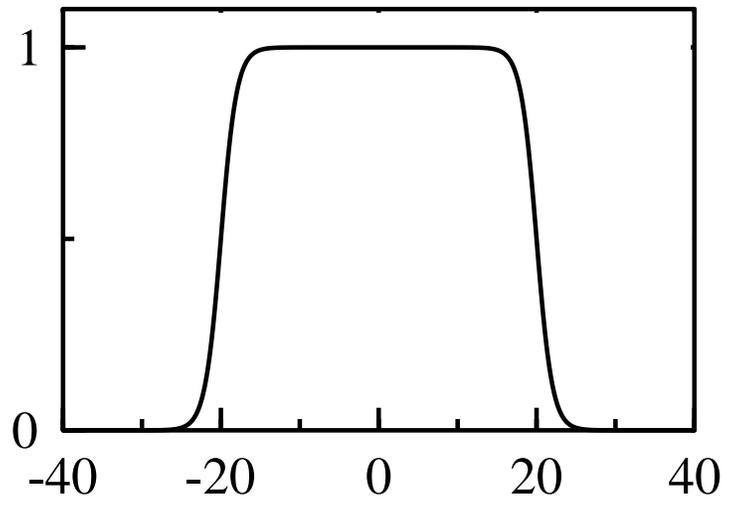
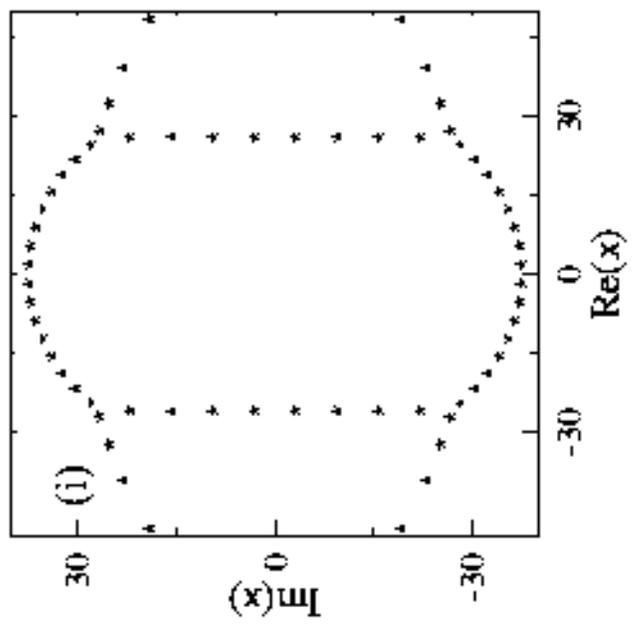
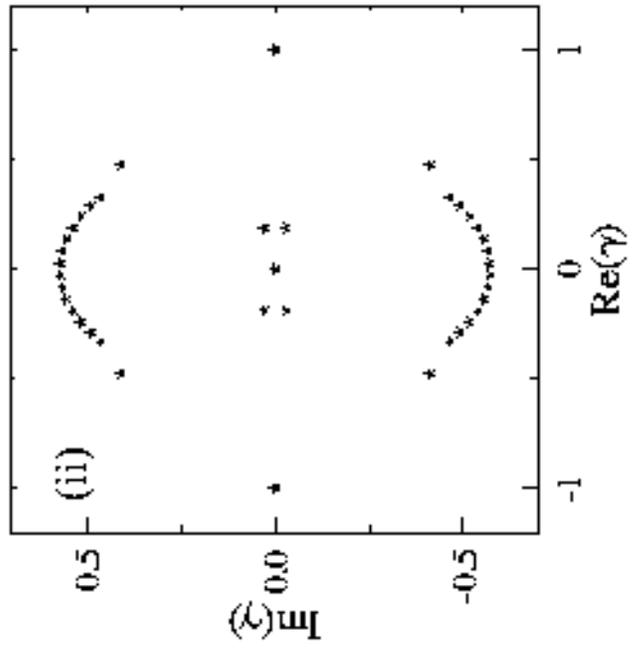
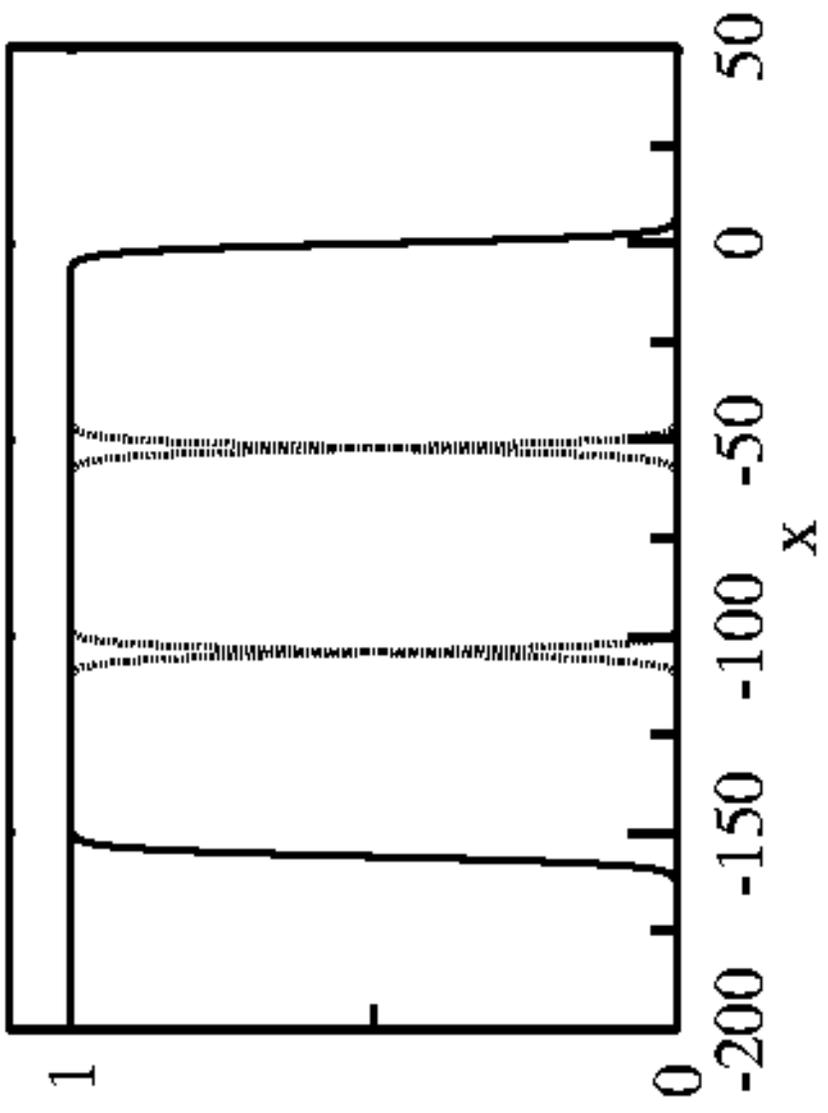
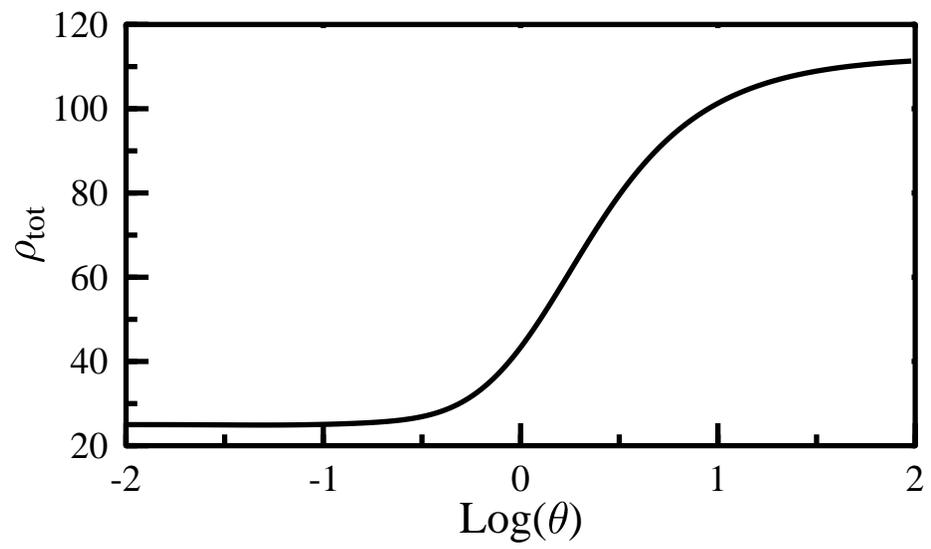
\end{document}